\documentclass{jetpl}
\usepackage{epsfig,amssymb}
\twocolumn

\lat

\title{
Testing the correlations between ultra-high-energy cosmic rays and BL Lac
type objects with HiRes stereoscopic data
}

\rtitle{Correlations of BL Lacs and HiRes cosmic rays}

\sodtitle{
Testing the correlations between ultra-high-energy cosmic rays and BL Lac
type objects with HiRes stereoscopic data 
}

\author{D.\,S.\,Gorbunov$^{\;a}$\/\thanks{e-mail: gorby@ms2.inr.ac.ru},
P.\,G.\,Tinyakov$^{\;b,a}$, I.\,I.\,Tkachev$^{\;c,a}$,
S.\,V.\,Troitsky$^{\;a}$}

\rauthor{D.\,S.\,Gorbunov et.\ al.}

\sodauthor{D.\,S.\,Gorbunov,
P.\,G.\,Tinyakov, I.\,I.\,Tkachev,
S.\,V.\,Troitsky}

\address{
$^a$ Institute for Nuclear Research of the Russian Academy of
Sciences,\\
60th October Anniversary Prospect 7a, 117312, Moscow, Russia;\\
$^b$
Service de Physique Th\'{e}orique, CP 225,
Universit\'{e} Libre de Bruxelles, B--1050, Brussels, Belgium;\\
$^c$ CERN Theory Division, CH-1211 Geneva 23, Switzerland.
}

\abstract{ Previously suggested correlations of BL Lac type objects with the
arrival directions of the ultra-high-energy cosmic ray primaries are tested by
making use of the HiRes stereoscopic data. The results of the study support
the conclusion that BL Lacs may be the cosmic ray sources and suggest the
presence of a small (a few percent) fraction of neutral primaries at
$E>10^{19}$~eV. }

\PACS{98.70.Sa, 98.54.Cm}

\begin{document}
\maketitle

\section{Motivation}

With the exception of a small number of highest-energy events, the
bulk of ultra-high energy cosmic rays (UHECR) is most naturally
explained by acceleration in active
galaxies~\cite{Ginzburg,Rachen:1992pg,Berezinsky:2002nc}.
This explanation is supported by clustering of UHECR observed in the AGASA
and Yakutsk datasets
\cite{Takeda:1999sg,Tinyakov:2001ic,Finley:2003ur}
\footnote{HiRes collaboration does not observe significant
clustering~\cite{Abbasi:2004dx,Abbasi:2004ib} but, at the present level of
statistics, this does not contradict to the previous
results~\cite{Yoshiguchi:2004np,Kachelriess:2004pc}.}
and by correlations between arrival directions of UHECR and
a particular class of active galactic nuclei, the BL Lacertae objects (BL
Lacs) \cite{BL1,BL:GMF,BL2} (see also Ref.~\cite{Uryson1}).

The correlations with BL Lacs were observed in AGASA and Yakutsk datasets.
To test them, {\em independent} datasets are needed. Such a dataset has
been recently published by the HiRes collaboration~\cite{Abbasi:2004ib}. It
consists of 271 events with reconstructed energy $E> 10^{19}$~eV which were
observed in the stereoscopic mode. The energies of the events are not
published; this makes reconstruction~\cite{BL:GMF} of arrival directions in
the Galactic magnetic field impossible. Despite the small exposure of HiRes
in the stereo mode as compared to AGASA, the excellent angular resolution
of $0.6^\circ$ makes HiRes data quite competitive (or even superior) in
testing correlations with BL Lacs under the assumption of neutral primary
particles. This test is the purpose of the present work.  More precisely,
we take three previously studied sets of BL Lacs without any further
selection and test them for correlations with the entire set of 271 HiRes
events.

\section{Fluctuations of correlation signal}
\label{signal}

At the current low statistics the correlation signal is not expected to be
robust \cite{Tinyakov:2001ic}. To illustrate this important point consider
the following example. Assume that a given set of sources in a given
exposure time produces 5 events in average. Let the average number of rays
coincident with sources by chance be 2. The {\em average} number of
correlating rays observed in such an experiment would therefore be 7. Both
the real events from sources and background chance coincidences are Poisson
distributed. Thus, in a particular realization the number of correlating
rays would vary roughly within $7\pm\sqrt{7}$. It is therefore likely that
two identical experiments will see 10 and 4 correlating rays,
respectively. The first one would see a strong correlation signal (10
events at 2 expected with the hypothesis of no correlations; probability
$p\sim 5\times 10^{-5}$), the second one would see a signal compatible with
the uniform background (4 events at 2 expected, $p\sim 14\%$).

It is worth stressing that, in this situation, the final conclusion
would be that the {\em correlation exists}, because the first of the
two experiments is incompatible with uniform background (while both
are compatible with the signal). Thus, at present small statistics one
{\em should} expect that some of the experiments see no
correlations. Such fluctuations indeed exist, e.g., in the
autocorrelation signal \cite{Takeda:1999sg,Abbasi:2004ib}.

\section{BL Lac samples}
\label{catalogs}

There were three samples of BL Lacs discussed in the literature, all
three drown from the catalog of quasars and active galactic nuclei
\cite{Veron2003}:

(1) The set of 22 most powerful BL Lacs, Ref. \cite{BL1}.
   This set was found by adjustment of cuts on magnitude and radio flux in
   the catalog of confirmed BL Lacs using the requirement of the best
   correlation signal with the most clustered subset of AGASA and Yakutsk
   cosmic ray events. Neutral primaries were assumed.

(2) The set of 14 BL Lacs which contains potentially $\gamma$-ray-loud
   objects, Ref. \cite{BL2}. This subset was obtained by the positional
   cross-correlation of confirmed BL Lacs with the catalog of EGRET sources
   \cite{3EG}. Both assumptions of neutral and charged primary particles
were
   tested using the most clustered subset of AGASA and Yakutsk events.

(3) The set of all confirmed BL Lacs with the single cut on
visual magnitude, ${\rm mag} < 18$, Ref. \cite{BL:GMF}. This set of BL
Lacs
contains 156 objects. Both neutral and charged primaries were tested using
   the whole available set of AGASA events, $E > 4\times 10^{19}~ {\rm
   eV}$. Better correlations were found under the assumption of positively
   charged primary particles.

Note that the subsets (1-3) are not independent. BL Lacs which correlate with
the UHECR in the set (1) are also present in the set (2). The cut on
magnitude, ${\rm mag} < 18$, being the single cut for the set (3), is also
imposed in the set (1) among other cuts.

\section{Procedure}

Our analysis is based on the calculation of the angular correlation
function by means of the algorithm described in Ref.~\cite{BL1}. The
statistical significance of correlation is estimated by testing the
hypothesis that the UHECR and BL Lacs are uncorrelated. This is done as
follows. For a given set of sources and the angle $\delta$, we count the
number of pairs ``cosmic ray -- source'' separated by the angular distance
less or equal $\delta$, thus obtaining the ``data count''. We then replace
the real data by a randomly generated Monte-Carlo set of cosmic rays and
calculate the number of pairs in the same way, thus obtaining the
``Monte-Carlo count''. We repeat the latter procedure many times noting
cases when the Monte-Carlo count equals or exceeds the data count. The
number of such cases divided by the total number of tries gives the
probability $P(\delta)$ which characterizes the significance of
correlations at a given angular scale $\delta$. The smaller is this
probability, the stronger (more significant) is the correlation.

The Monte-Carlo events are drown from the isotropic distribution according
to the acceptance of a given UHECR detector. For the HiRes experiment in
the stereo mode we use the zenith-angle and azimuth-angle distributions
given in Ref.~\cite{stereo-ICRC} and the sidereal time distribution
published in Ref.~\cite{H-exp}.

The probability $P(\delta)$ depends on the choice of $\delta$. The optimal
value of $\delta$ is clearly the one where the {\em expected} signal is
strongest (e.g., expected $P_{\rm th}(\delta)$ is lowest). In the case of
point sources and neutral primary particles the shape of $P_{\rm th}(\delta)$
can be calculated by the following Monte-Carlo simulation. We start by
generating a cosmic ray set which is artificially correlated with BL
Lacs. This is achieved by replacing a given (small) number of events in a
random set by the events correlated with randomly chosen BL Lacs. To generate
an event correlated with the BL Lac we assume the 2-dimensional Gaussian
distribution normalized in such a way that the circle of the radius
$0.6^\circ$ around the BL Lac contains 68\% of events.  The resulting
correlated set is then treated as the real data, i.e., the probability
$P(\delta)$ is calculated. The bin-by-bin median values of the probabilities
obtained for many different sets give $P_{\rm th}(\delta)$.  In the real data,
following Ref. \cite{Tinyakov:2003bi}, we measure correlations at that value
of $\delta$ where $P_{\rm th}(\delta)$ is minimal.

\section{Results}

\begin{figure}
\begin{center}
\includegraphics[width=9cm]{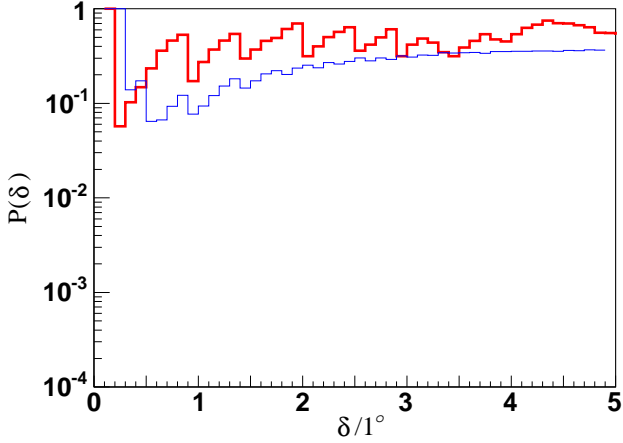}
\caption{FIG. 1. $P(\delta)$ for the set of 22 BL Lacs and HiRes stereo
events. The thick line shows data, the thin line shows $P_{\rm th}(\delta)$
obtained in the Monte-Carlo simulations in which 2 events are from the BL Lac
sources.}
\label{fig:22}
\end{center}
\end{figure}

\begin{figure}
\begin{center}
\includegraphics[width=9cm]{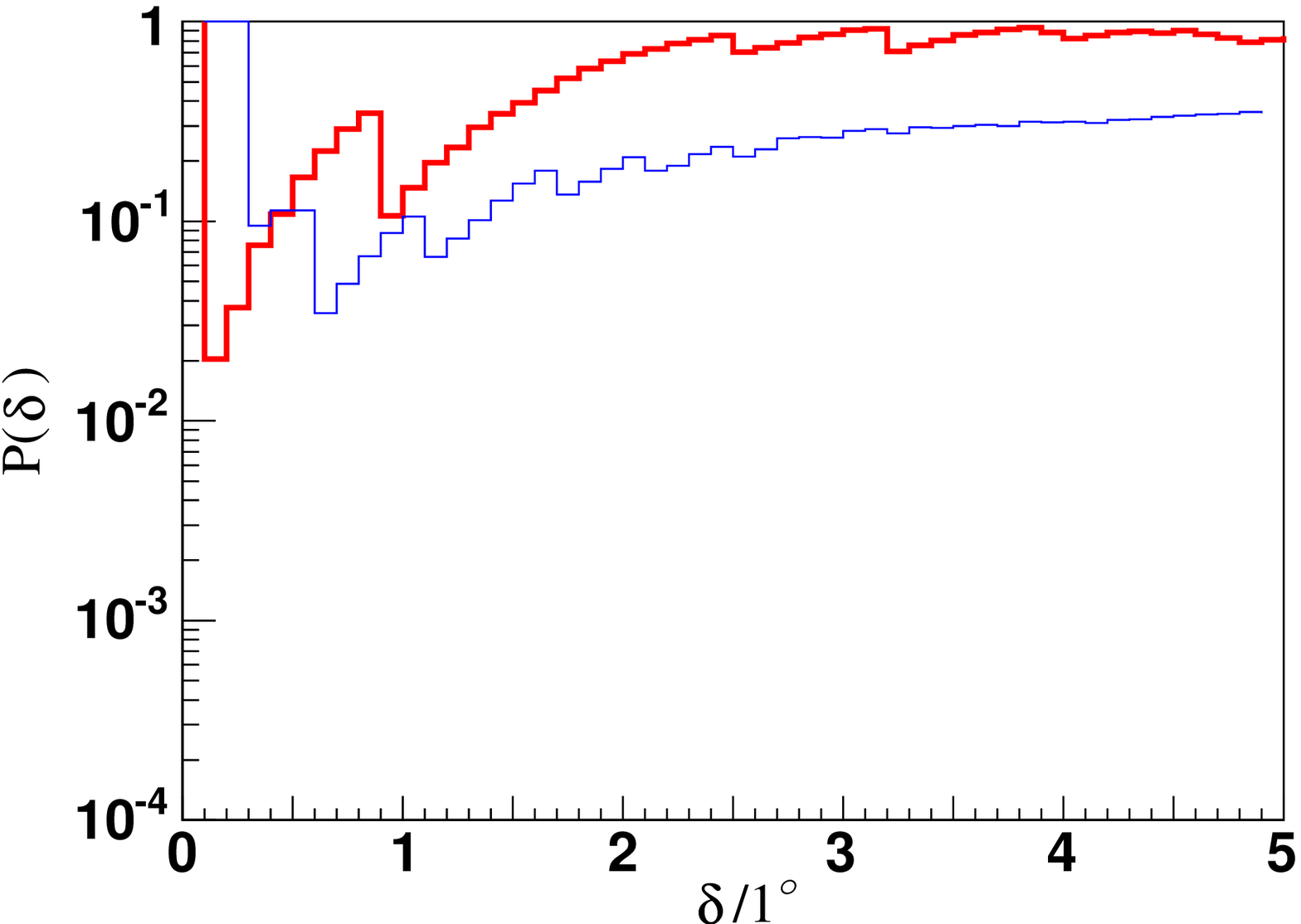}
\caption{FIG. 2. $P(\delta)$ for the set of 14 BL Lacs and HiRes stereo
events. The thick line shows data, the thin line shows $P_{\rm th}(\delta)$
obtained in the Monte-Carlo simulations in which 2 events are from the BL Lac
sources. }
\label{fig:14}
\end{center}
\end{figure}

\begin{figure}
\begin{center}
\includegraphics[width=9cm]{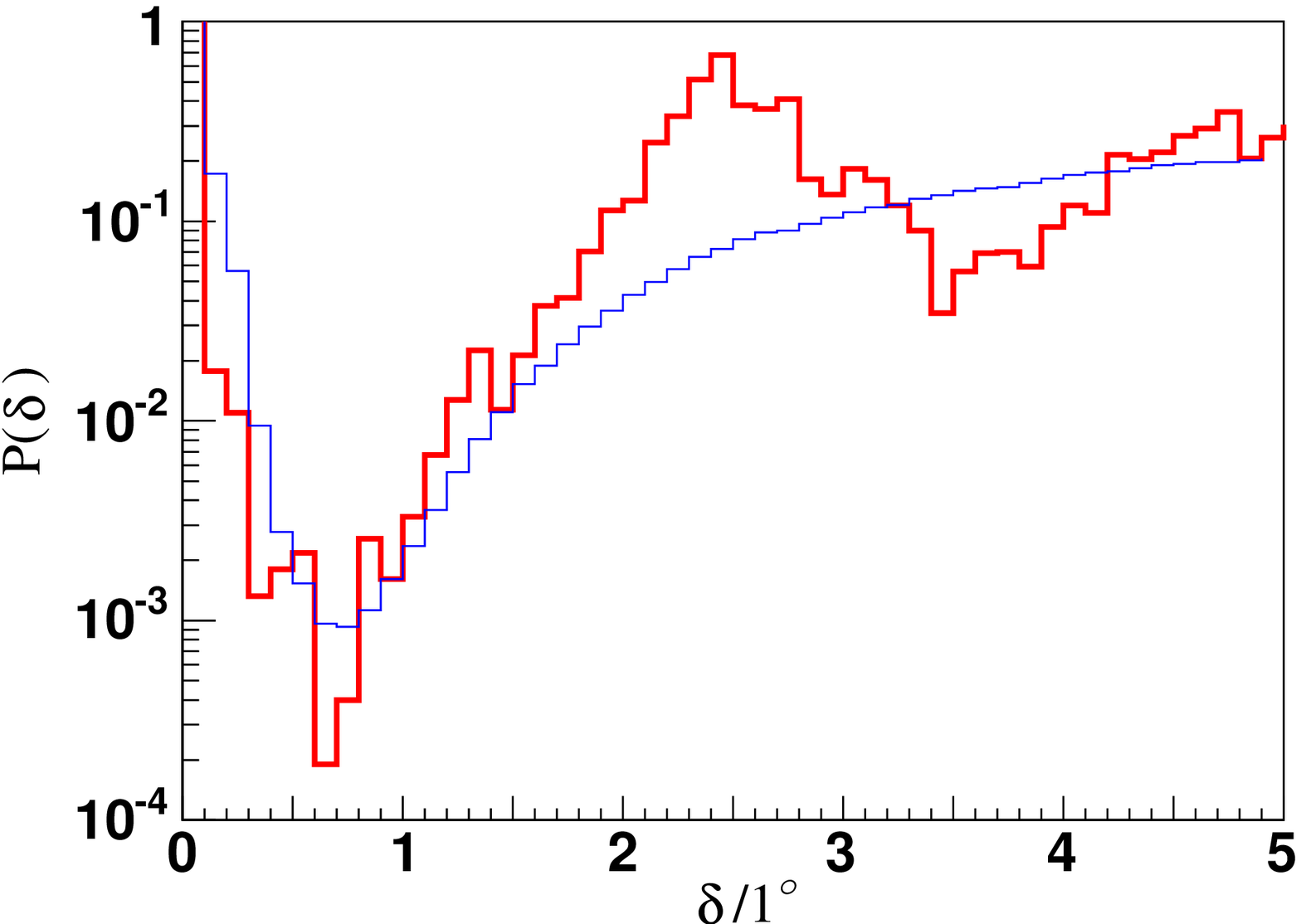}
\caption{FIG. 3. $P(\delta)$ for the set of 156 BL Lacs and HiRes stereo
events. The thick line shows data, the thin line shows $P_{\rm th}(\delta)$
obtained in the Monte-Carlo simulations in which 9 events are from the BL Lac
sources.}
\label{fig:156}
\end{center}
\end{figure}

In Figures \ref{fig:22} - \ref{fig:156} we present the expected (thin line)
and observed (thick line) $P(\delta)$ for the HiRes dataset of UHECR
and three samples of BL Lac objects described in
Section~\ref{catalogs}. There are no significant correlations with the
subsets of 22 and 14 BL Lacs. In both cases only two BL Lacs contribute to
data counts within 1 degree. In the subset of 22 BL Lacs those are TXS 
0751+485 and OT 465 (the latter object was among 5 BL Lacs which
contribute to correlations in Ref.~\cite{BL1}). In the subset of 14 BL
Lacs two contributing objects are PKS 1604+159 and 1ES 1959+650. It is
worth mentioning that 1ES 1959+650 is a confirmed TeV source and
contributes to correlations in Ref~\cite{BL2}.

The set of 156 BL Lacs with ${\rm mag}<18$ shows rather strong
correlation. We list in Table~1 the cosmic rays and
BL Lac's separated by less than $1^\circ$.
\begin{table}
\begin{tabular}{|c|c||c|c|}
\hline
\multicolumn{2}{|c||}{Cosmic ray}&
\multicolumn{2}{c|}{BL Lac}\\
\hline
$\alpha $, deg. & $\delta $, deg.&name & $z$\\
\hline
 17.8 & $-$12.5 &RBS 0161      & 0.234  \\
 48.5 &   5.8 &RX J03143+0620& ?      \\
118.7 &  48.1 &TXS 0751+485  & ?      \\
123.8 &  57.0 &RX J08163+5739& ?      \\
137.2 &  33.5 &Ton 1015      & 0.354  \\
162.6 &  49.2 &MS 10507+4946 & 0.140  \\
169.3 &  25.9 &RX J11176+2548& 0.360  \\
209.9 &  59.7 &RX J13598+5911& ?      \\
226.5 &  56.5 &SBS 1508+561  & ?      \\
229.0 &  56.4 &SBS 1508+561  & ?      \\
253.7 &  39.8 &RGB J1652+403 & ?      \\
265.3 &  46.7 &OT 465        & ?      \\
300.2 &  65.1 &1ES 1959+650  & 0.047  \\
\hline
\end{tabular}
\caption{Table 1. Cosmic rays and BL Lac's from the sample of 156 objects
separated by less than $1^\circ$. Note that SBS 1508+561 correlates with two
rays.}
\end{table}
The expected probability $P_{\rm th}(\delta)$ has minimum at $\delta=
0.8$. At this value of $\delta$ the real data give the probability $P(0.8) =
4\times 10^{-4}$.  The data count is 11, while the Monte-Carlo expectation
due to random uncorrelated background is $\sim 3$. This corresponds to about
8 events from sources in the overall HiRes data-sample of 271 events. Thus,
a $\sim 3\%$ fraction of neutral particles in the total cosmic ray flux may
be sufficient to explain this correlation. This is consistent with a
fraction of clustered events observed by AGASA at these
energies~\cite{AGASA-clust}.

\section{Conclusions}

We have assumed --- as a ``null-hypothesis'' --- that BL Lacs and the
new HiRes stereo data are uncorrelated. We have tested this hypothesis
for 3 particular subsets of BL Lacs studied previously and found an
excess of pairs ``BL Lac -- cosmic ray'' in one of them, which would
occur with the probability $4\times 10^{-4}$ for a random set of
UHECR. If these sets were independent, the final significance would be
the best probability found multiplied by the number of tries. Since
the samples are not independent, the penalty factor is smaller. In any
case, the null-hypothesis is rejected with the probability $\sim
10^{-3}$. Our analysis thus confirms the association of UHECR with BL
Lacs.

Only one of the three examined subsets of BL Lacs --- the largest one
--- shows significant correlations. This may indicate that those BL
Lacs which are sources, have similar and small luminosity in UHECR.
This conclusion is also consistent with the statistics of the
clustering of UHECR events, which suggests few hundred sources
currently contributing to the UHECR flux, see
Refs.\cite{Dub,Blasi:2003vx,Kachelriess:2004pc}.

The observation of correlations at angles which are much smaller than the
typical deflection of a charged UHECR particle in the Galactic magnetic
field suggests that (at least) a few percent of UHECR are neutral.

The authors are indebted to K.~Belov, V.~Berezinsky, M.~Libanov, A.~Neronov,
D.~Semikoz and T.~Weiler for interesting discussions and useful comments.
This work was supported in part by the INTAS grant 03-51-5112 (D.G., P.T.\ and
S.T.), by the RFBR grant 02-02-17398 (D.G.\ and S.T.), by the grants of the
President of the Russian Federation NS-2184.2003.2, MK-2788.2003.02 (D.G.),
MK-1084.2003.02 (S.T.), by the grants of the Russian Science Support
Foundation (D.G.\ and S.T.), by the fellowships of the "Dynasty" foundation
(awarded by the Scientific board of ICFPM; D.G.\ and S.T.) and by the Swiss
Science Foundation, grant 20-67958.02 (P.T.).

\end{document}